\documentstyle[astrobib,psfig,epsfig]{mn-ab}

\def\Mpc{\ifmmode {\, h^{-1} \, {\rm Mpc}}
\else {$h^{-1}\,$ Mpc}\fi}

\def\s8{{\sigma_8}}

\def\deg{^\circ} 
\def\ltsima{$\; \buildrel < \over \sim \;$}
\def\simlt{\lower.5ex\hbox{\ltsima}} 
\def\gtsima{$\; \buildrel > \over \sim \;$} 
\def\simgt{\lower.5ex\hbox{\gtsima}} 
\def\gsim{\lower.5ex\hbox{\gtsima}}

\begin{document}
\title[]{Constraints on Cosmological Anisotropy out to $z=1$ from 
         Supernovae Ia}
\author[Kolatt \& Lahav ] 
{Tsafrir S. Kolatt$^{1}$ and  Ofer Lahav$^{2,1}$\\
$^1$ Racah Institute of Physics, The Hebrew University, Jerusalem 91904, Israel\\
$^2$ Institute of Astronomy, Madingley Rd., CB3 0HA, Cambridge, UK \\
}

\maketitle

\begin{abstract}
A combined sample of 79 high and low redshift supernovae Ia (SNe) is used to
set constraints on the degree of anisotropy in the Universe out to $z\simeq1$.
First we derive the global most probable values of matter density $\Omega_M $,
the cosmological constant $\Omega_\Lambda $, and the Hubble constant $H_0$, and
find them to be consistent with the published results from the  two data sets of
Riess et al. 1998 (R98) and Perlmutter et al. 1999 (P99).  We then examine the
Hubble diagram (HD, i.e., the luminosity-redshift relation) in different
directions on the sky by utilising spherical harmonic expansion.  In
particular, via the analysis of the dipole anisotropy, we divide the sky into
the two hemispheres that yield the most discrepant of the three cosmological
parameters, and the scatter $\chi^2_{\rm HD}$ in each case.  The most
discrepant values roughly move along the locus $-4\Omega_M +3 \Omega_{\Lambda}
= 1$ (cf. P99), but by no more than $\Delta \approx 2.5$ along this line.
For a  
perfect FRW universe, Monte Carlo realizations that mimic the current set of
SNe yield values higher than the measured $\Delta$ in $\sim 1/5$ of the cases.
We discuss implications for the validity of the Cosmological
Principle, and 
possible calibration problems in the SNe data sets.

\end{abstract}

\section{Introduction}
\label{sec:intro}

The validity of the Cosmological Principle and the isotropy it implies gained
much credibility in recent years.  The small fluctuations in the CMB ($\Delta
T/T \sim 10^{-5}$ 
on angular scale $\sim 10^\circ$) provide the strongest evidence that the
universe can be well approximated by the FRW metric on scales larger than $\sim
1000 \Mpc $ (e.g., Peebles 1993; Wu, Lahav, \& Rees 1999)

On smaller scales
($\sim 100 \Mpc$) bulk flows of the order $v/c \sim 10^{-3}$ indicate that this
isotropy breaks down.  This is also manifested by 
significant correlation functions of 
galaxies and clusters on large scales, 
and structures like the
Supergalactic Plane and the Great Attractor.  The transition scale to isotropy
and homogeneity is still poorly known, and so is the convergence of the
acceleration vector of the Local Group with respect to the CMB.  It is
therefore important to quantify the degree of homogeneity and isotropy as
function of scale. Traditionally this was done by searching for anisotropy in
the distribution of radio sources and background radiations 
\cite{nan-cai:96,evans:92,webster:76}.
Several
new methods have been suggested to test isotropy and homogeneity on redshift
scales of $z \approx 0.1-5$, 
such as
measurements of {\it in situ} CMB temperature \cite{songaila:94}, 
the derivation of an
independent rest frame from multiple image lens systems 
\cite{kochanek-kolatt-msb}, and Faraday rotation signature due to anisotropic magnetic
field 
\cite{kronberg:76,vallee:90,nodland-ralston:97}.

The recent use of SNe as distance indicators 
\cite{phillips:93,perlmutter:95,RPK:96}
opened a new opportunity for
accurate measurements of anisotropy on cosmological scales that
previously have not been accessible. So far the SNe have been used
in order to constrain the Hubble constant $H_0$ 
from a nearby sample and
combinations of the matter density $\Omega_M$
and the cosmological constant $\Omega_{\Lambda}$ utilizing 
SNe at moderate ($\ga0.3$) and high ($\sim1$) redshifts.
In the future, SNe samples over a wider  redshift range 
will provide separate estimates for the two parameters.
It is important to establish the `universality' of the
measurements of cosmological parameters from SN, as
they are commonly used in joint analysis with other probes
such as the CMB, cluster abundance and peculiar velocities 
\cite{efsth:99,efst-bridle:99,bridle:99,bridle:00,tegmark:99}. 

Assuming a FRW cosmology,
a forth  measure can be deduced from the `Hubble diagram'
(HD; i.e., the luminosity -- redshift relation),
the $\chi^2_{HD}$ measure for the best fit model.
For a perfect distance indicator
this measure  indicates
deviations of the local potential (i.e., at the location
of the SN) from a pure FRW geometry.
However, in the real universe 
the deviations can also be due to other sources:

\begin{itemize}
\item Intrinsic (astrophysical) scatter in the SN luminosity-light curve
relation.

\item
Scatter due to the location of  the SN 
within the host galaxy  \& the
galaxy type.

\item
Scatter due to dust absorption in the host galaxy,
in the intergalactic medium and in our Galaxy.

\item
Gravitational  lensing along the l.o.s. to the SN
(e.g., an overdensity along the l.o.s. will enhance the apparent 
luminosity of a SN).

\end{itemize}

Here we explicitly assume that there is no
evolution with redshift in the luminosity-light curve relation.
Fortunately, most of the abovementioned effects are  on the scale
of the host galaxy, so with large enough sample they would be averaged
out in the calculation of large scale anisotropies.
On the other hand, 
one should worry about `anisotropies' which are    
simply due to poor matching of different data sets 
that sample different portions of the sky,
or large angular effects due to Galactic extinction.

We also note that
some of these effects above might be correlated with other measurements,
e.g. if the scatter $\chi^2_{HD}$  detected in
SN Hubble diagram is affected
by fluctuations in the potential, then it would be correlated with
Integrated SW (or Rees-Schiama) effect in the CMB fluctuations.

The outline of this paper is as follows, in \S\ref{sec:data} we present
the unified data set we will be using for the isotropy analysis. The
results for cosmological parameters from the entire sample are presented
in \S\ref{sec:unified}, the anisotropy measurement is discussed in
\S\ref{sec:anisotropy}, and put in a probabilistic context in
\S\ref{sec:degree}.  We conclude our results in \S\ref{sec:discuss}.

\section{The Unified Data set}
\label{sec:data}

An ideal data set of SNe for the goals we have put forward in the
introduction would be a whole-sky homogeneous coverage at various
redshifts of SNe. Since such an optimal set does not exist, the closest
data set would be the amalgamation of the two existing, published data
sets.

We unify the samples of the Supernova Cosmology Project 
(SCP) \cite{perlmutter:99}
and that of the High-z Supernova search team (HZS) \cite{riess:98}.
These include also the data from low redshift of the Cal\'an-Tololo
survey \cite{hamuy:96}.
The two groups have different strategy and different nomenclature for
the minimization problem by which the cosmological parameters are
derived.
We have brought the SCP data to comply with the language of the HZS
team 

\begin{figure}
\includegraphics{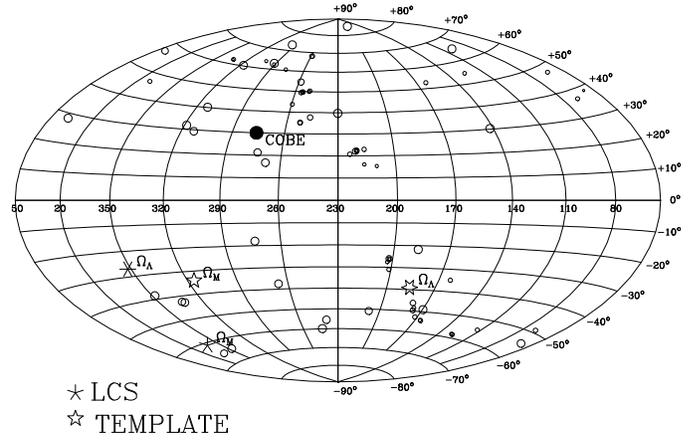}
\vskip5.5truecm
\caption{
The sky distribution in Galactic coordinates of the 79 SNe composing the
unified sample. The point size is proportional to $(1+z)^{-1}$ of the SNe.
Also shown are the (positive) directions that maximize the
$\Omega_M$ and $\Omega_\Lambda$ dipoles using the two methods (cf.
\S\ref{sec:anisotropy}),
and the CMB dipole direction in the Local Group rest-frame
as measured by COBE.}
\label{fig:SN_sample}
\end{figure}

For each SNe we list its (i) $cz$ in the CMB frame,
(ii) the distance modulus  $\mu = m_B^{eff}-M_B^{fiducial}$,
(iii) errors for these two quantities, (iv) Galactic $l$ and $b$.
For the SCP data the fiducial magnitude, $M_B^{fiducial}$ (cf. P99), 
is obtained by comparison of the 18 
overlapping low redshift SNe 
as analysed by the two groups, and
equating the distance modulus of R98 (table 10) to $m_B^{\rm corr} $
of P99 (table 2).
This procedure is repeated twice, since Riess et al. provide two ways to
calculate the distance moduli, ``Multi Light Curve Shapes'' (LCS) and
``Template". Errors are taken from the tables and  a least square 
minimization is performed in order to obtain the two best fit values of
$M_B^{fiducial}$ of P99
(and to recover the Hubble constant dependence they omitted
in their calculation).
The two values are $M_B^{fiducial}+5\log H_0=-19.322, -19.453$ with
$\chi^2/d.o.f$ of 0.952 and 0.763
for the LCS method and the TEMPLATE method respectively.
The value of $M_B^{fiducial}$ is degenerated with $H_0$, 
so different $H_0$ calibrations in the two samples get ``absorbed" in
the value for $M_B^{fiducial}$.
The unified sample consists of 79 SNe altogether, after the exclusion of
6 SNe from P99 (taking their ``model C" version) and including the
snap-shot survey from R98 along with 1997ck.
Figure \ref{fig:SN_sample} shows the SNe distribution in Galactic
coordinates. 

The sky coverage is clearly inhomogeneous: the SNe deficiency near the
Galactic plane is evident and the clustering of  a few of the
observed SNe due to the detection procedure is clear.

\section{Cosmological parameters from the unified sample}
\label{sec:unified}

We follow the statistical analysis as described in R98 and obtain best
values for $H_0$ and probability contours in the ($\Omega_M,
\Omega_\Lambda$) plane after integration (i.e. 
marginalization) over all $H_0$ values and taking
into account only physical regions in that plane.

\begin{figure*}
{\epsfxsize=2.7 in \epsfbox{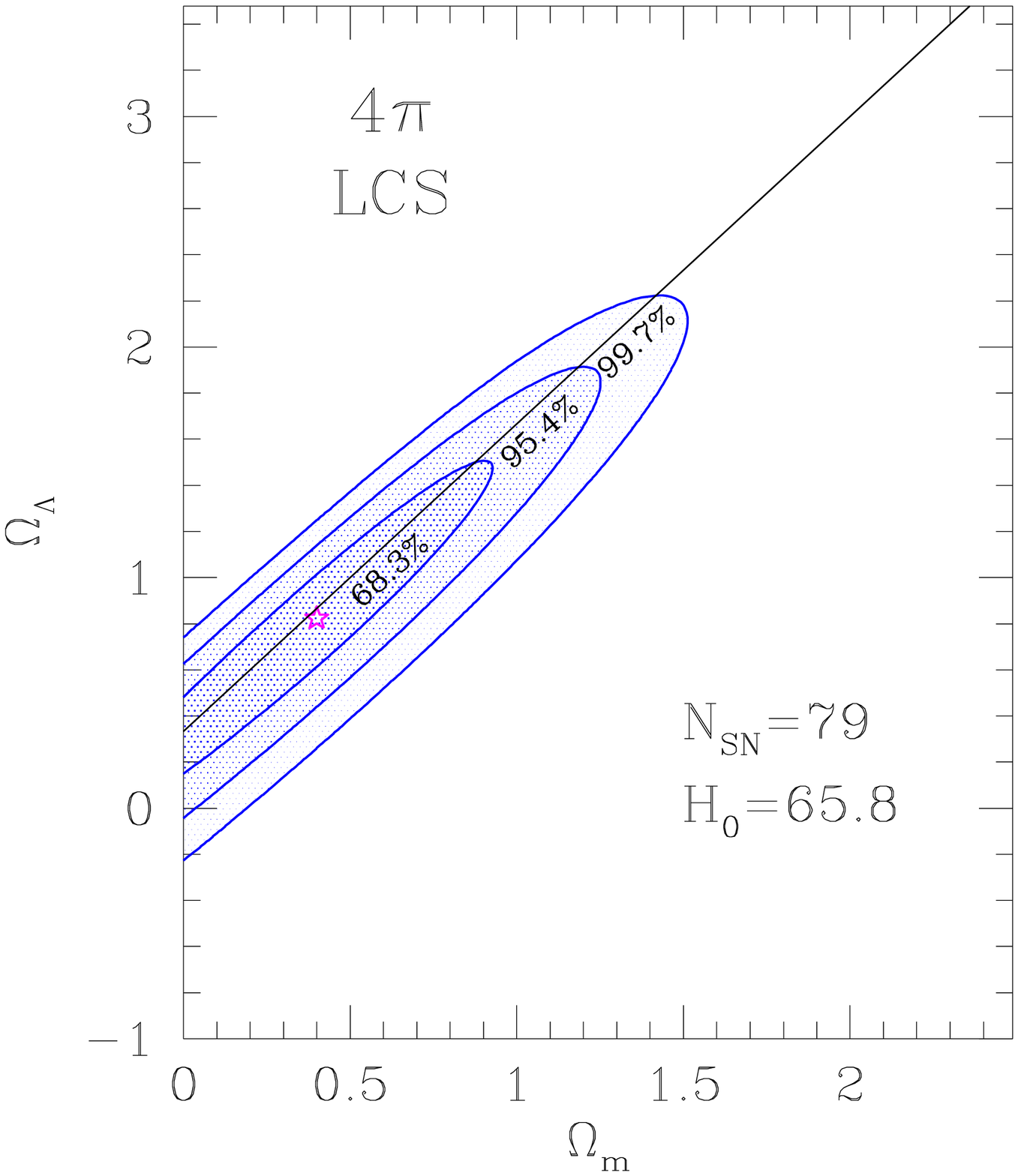}}
{\epsfxsize=2.7 in \epsfbox{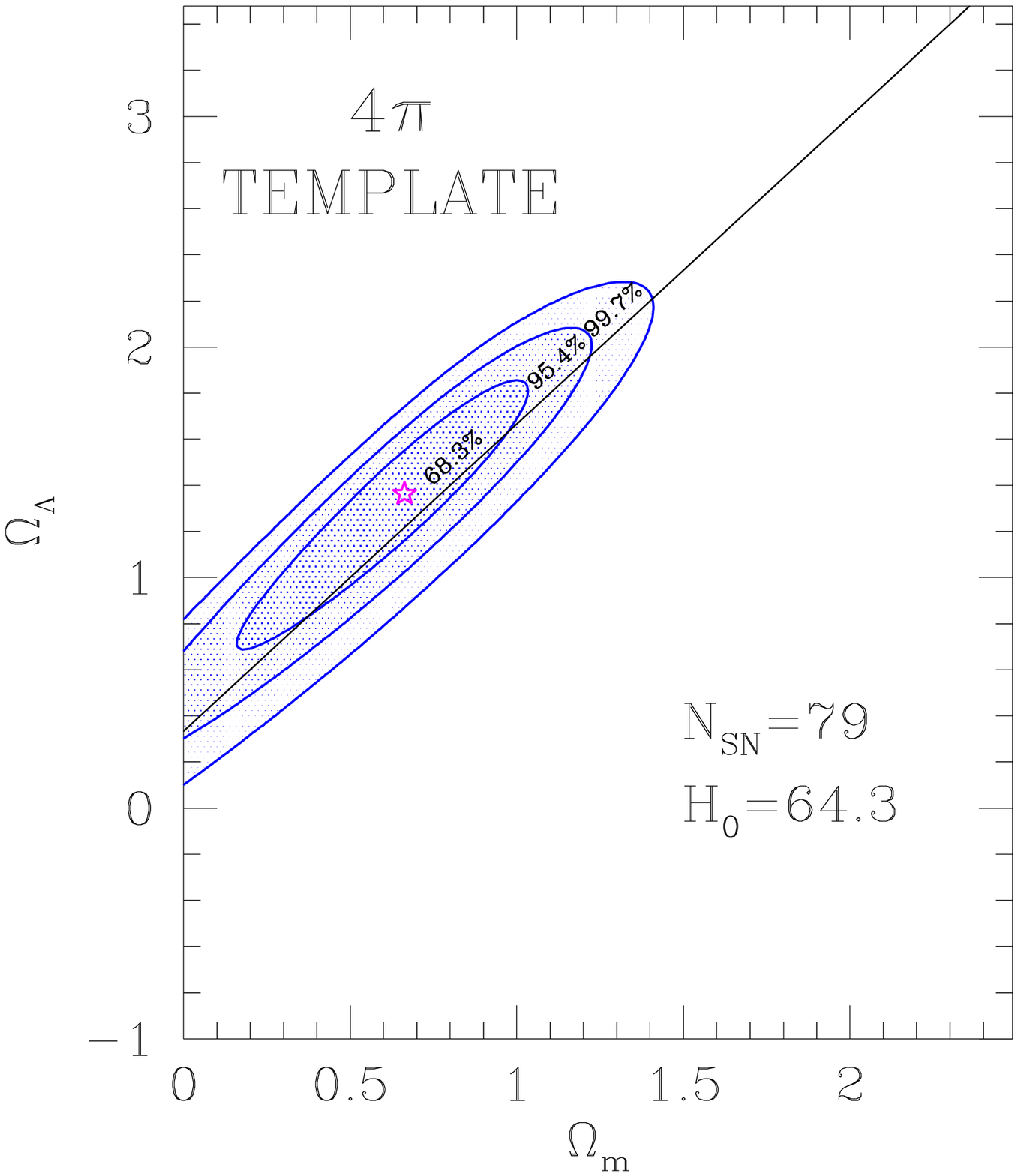}}
%\vskip-0.5truecm
\caption{ 
Confidence regions drawn from 79 SNe of the unified sample (see text),
using the LCS method (left) and the TEMPLATE method (right).
The best-fit parameters are marked by a star and the quoted best-fit line
from P99 ($0.8\Omega_M-0.6\Omega_\Lambda = -0.2$) is shown for reference 
as the diagonal line across the figure. Non-physical regions in the
$(\Omega_M,\Omega_\Lambda)$ plane are excluded.
}
\label{fig:xlikeA}
\end{figure*}

P99 include the error due to redshift measurements and peculiar
velocities in their magnitude errors, for R98 we followed their
procedure, set $\sigma_v=200$ km s$^{-1}$ for SNe of $z<0.5$ and
$\sigma_v=2500$ km s$^{-1}$ for SNe with $z\ge0.5$, and translated to the
distance modulus, $\mu$, units
according to the assumed cosmological model in the likelihood function.
Figure \ref{fig:xlikeA} show the results of the likelihood analysis.
The maxima of the likelihood functions are obtained for 
($\Omega_M, \Omega_\Lambda$) values of ($0.40,0.82$) and ($0.66,1.36$) 
for the LCS and TEMPLATE method respectively. The contour lines
correspond to the $68.3\%$, $95.4\%$, and $99.7\%$ confidence levels.

\section{Anisotropy measurement}
\label{sec:anisotropy}

The natural expansion for anisotropy detection is in spherical
harmonics. The current data are too sparse to 
allow analysis in redshift shells. 

We expand the four two-dimensional parameter `fields'
(for $\Omega_M$, $\Omega_\Lambda$, $H_0$, and
$\chi^2_{HD}$)
in spherical harmonics. 
If the isotropy assumption is valid we expect deviations from
the average value to be due to noise, and the angular power
spectrum should likewise reflect it. This is unless foreground effects
alter the signal significantly. 

The operational way to calculate the expansion coefficients $a_{lm}$ is as
follows.

\begin{itemize}
\item
Build a random distribution of points (``mask") on the sphere.

\item
Assign four best-fit parameters to every point
based on minimization over all SNe within angular radius $\gamma_{min}$
about this grid point.

\item
Construct the four residual fields about the global mean, i.e., 
$\delta_F=(F-\langle F\rangle)/\langle F \rangle$, where $F$ is
$\Omega_M$, $\Omega_\Lambda$, $H_0$, or $\chi^2_{\rm HD}$.

\item
Expand the $\delta_F$ values as obtained at each grid point in Spherical 
Harmonics up to $l_{max} = \pi/\gamma_{min}$, i.e.
\begin{equation}
\delta_F(\theta,\phi) = \sum_{l=0}^{l=l_{\max}}
\sum_{m=-l}^{m=+l}a_l^mY_l^m \,.
\end{equation}

\end{itemize} 
In order to include more than $2$
SNe in each smoothing bin (at least two-parameter fit) we obtain
$\gamma\simeq 25\deg$, however the SNe are not distributed uniformly (cf.
Fig. \ref{fig:SN_sample}) and
thus a minimum angular resolution of $\sim 60\deg$ is imposed.
That means that 
for a whole sky coverage 
the highest significant multipole, $l$, is $l=3$.
There are, though regions that are more densely covered by SNe
data and therefore higher multipoles can be assessed as well but at a lower
signal-to-noise level.

In order to account for the Poisson noise contribution (and thus
to the angular power spectrum in quadrature), we run a set of $50$ random
``masks" and repeat the $a_{lm}$ calculation each time.  
For each set of $a_{lm}$ the power spectrum coefficients,
$C_l=(2l+1)^{-1}\sum_{m=-l}^{l}\vert a_{lm} \vert ^2$ are computed.

\begin{figure*}
{\epsfxsize=2.7 in \epsfbox{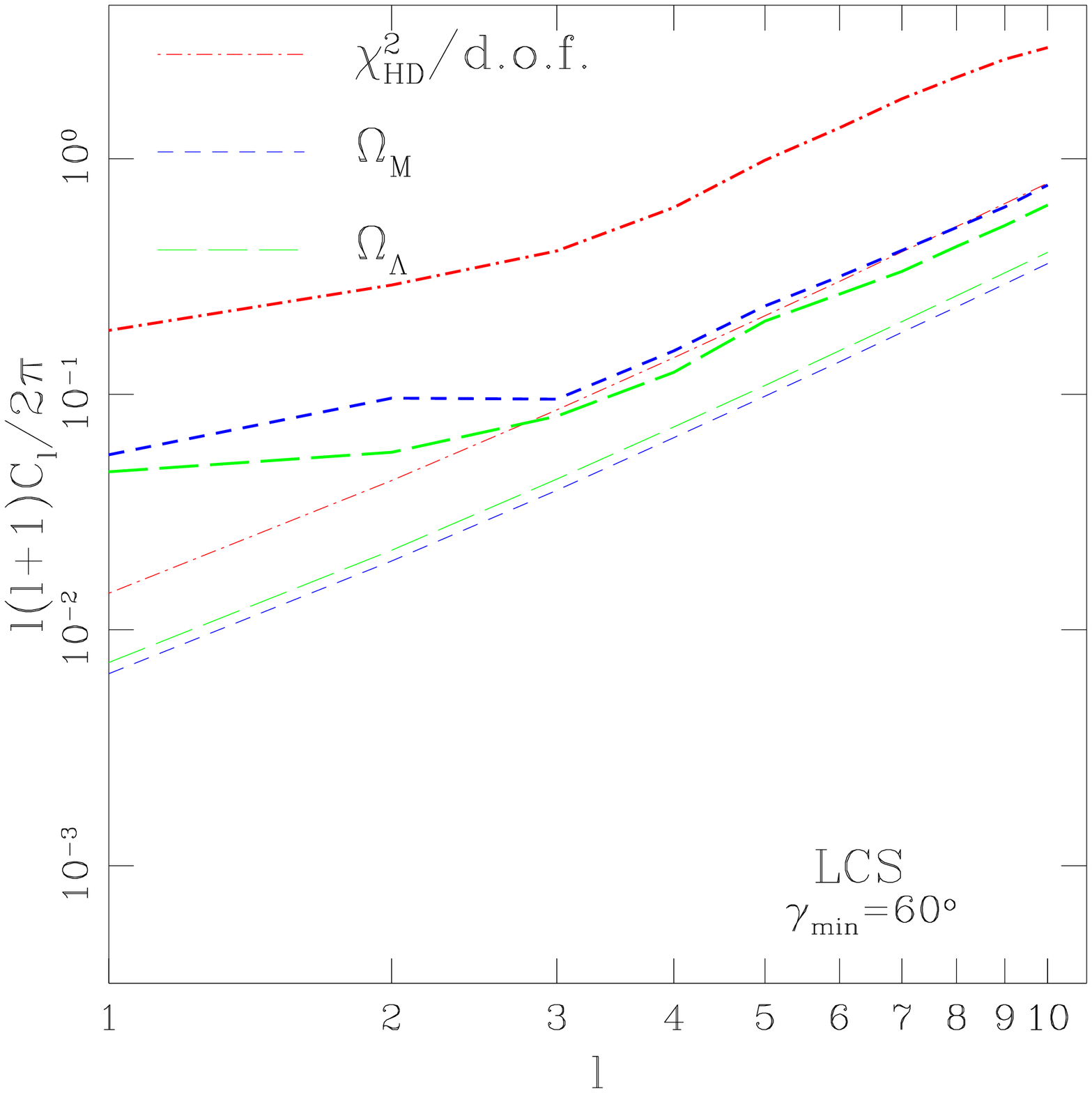}}
{\epsfxsize=2.7 in \epsfbox{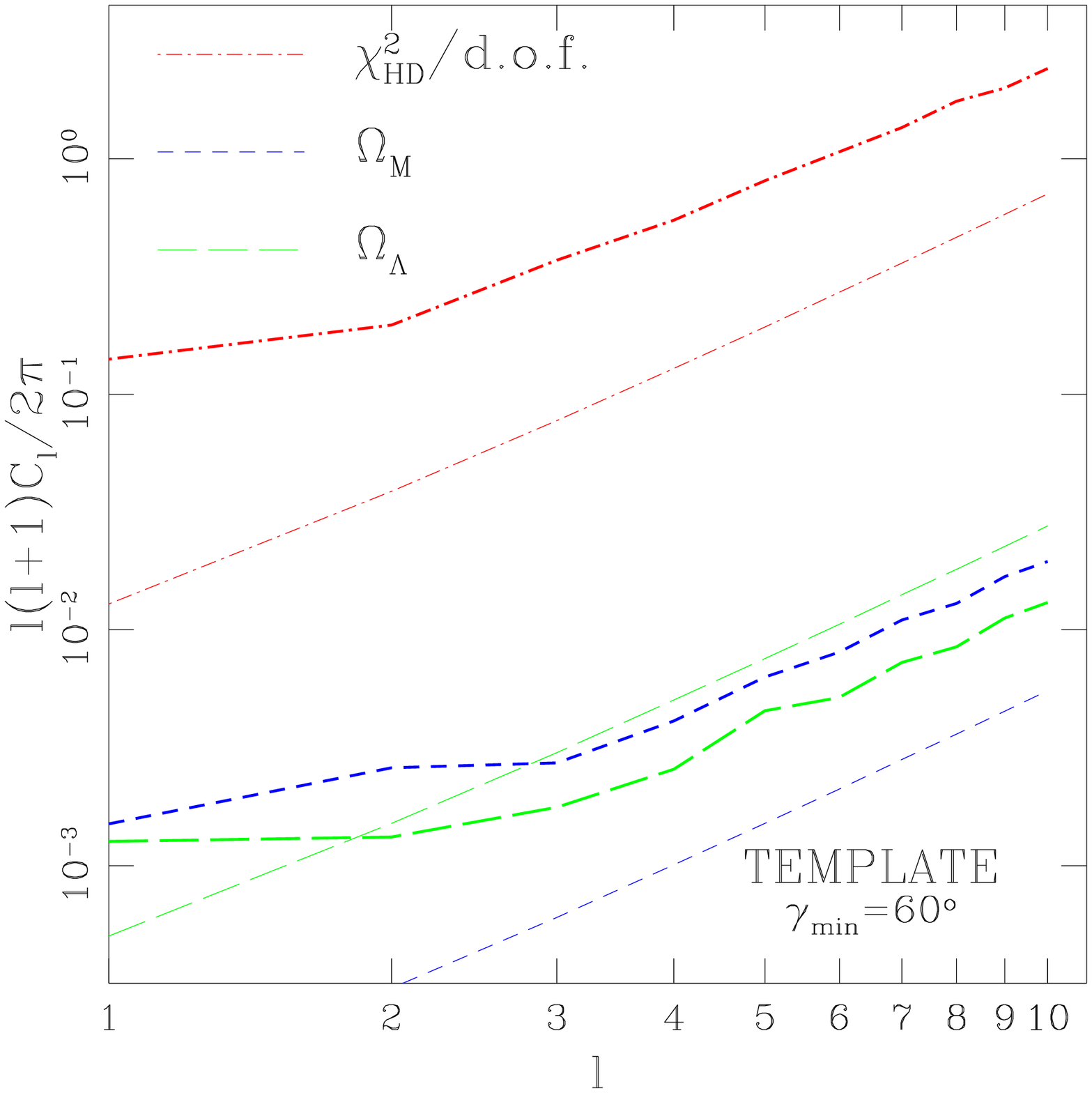}}
\caption{
Angular power spectrum coefficients for the two dimensional fields:
$\delta_{\Omega_M}$, $\delta_{\Omega_\Lambda}$, and 
$\delta_{\chi^2_{\rm HD}}$ along with the noise level of each field 
(straight weak lines). Shown are results from the
LCS method (right) and the TEMPLATE method (left).
} 
\label{fig:c_ls}
\end{figure*}

%\section{Dipole confidence levels}
The angular power spectrum of the $\delta_{H_0}$ is an order of
magnitude and more smaller than the noise level ($C_l/(2\pi) \simeq
5\times 10^{-5}$), in both methods.
Figure \ref{fig:c_ls} shows the angular power spectrum,
$l(l+1)C_l/(2\pi)$, for the other three
fields as calculated from 50 runs with different random mask points. 
The straight weaker lines show the noise level in each field.
The $\delta_\Omega$ fields in both methods show signals that
exceed the noise level for the dipole ($l=1$) and the quadrupole
($l=2$). Two factors contribute to the noise level, the discrete number
of SNe, and the scatter in the luminosity --- redshift relation.
The former is common to both methods (LCS and TEMPLATE) and therefore
the order of magnitude higher noise level for the $\delta_\Omega$ fields
in the LCS method must be due to the latter.
The TEMPLATE method seems to provide smaller errors and a better match
between the two data sets, as indicated by the lower $\chi^2$ level
of the fiducial magnitude calibration (cf. \S\ref{sec:data}).
The $\delta_{\chi^2_{\rm HD}}$ angular power spectrum is similar in shape
and magnitude in both methods, and lies an order of magnitude to a
factor $\sim5$ above its noise level. This may indicate there exists a
true dipole (or quadrupole) in this field.
From the first multipole of angular power spectrum alone, 
one cannot deduce what is the
dipole {\em direction}. We therefore turn to look for the direction by
other means.

We search for largest dipole in $\Omega_M$, $\Omega_\Lambda$, $H_0$, and
$\chi^2_{HD}$. This has been done in two ways : an actual search over the
sky, dividing the SN population in between two hemispheres, and
equivalently, 
by solving a maximization problem of the dipole term with respect to
$(\theta,\phi)$ using the computed $a_{lm}$ coefficients.
Both methods yield similar results.
We then calculate the confidence regions for each hemisphere
separately, and look for statistical consistency (overlapping contours).
Each test can be applied to each one of the four parameters.

\begin{figure*}
{\epsfxsize=2.7 in \epsfbox{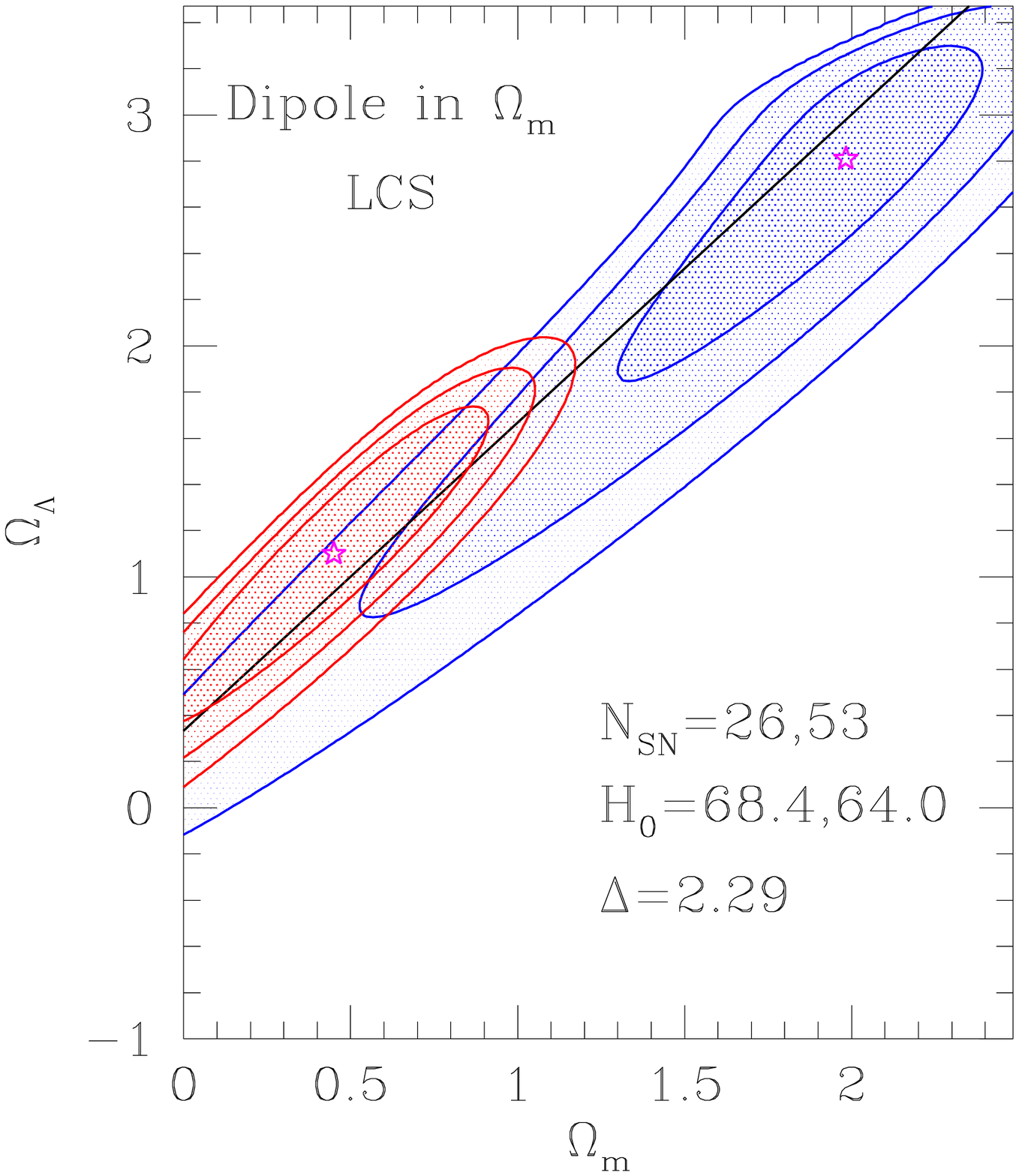}}
{\epsfxsize=2.7 in \epsfbox{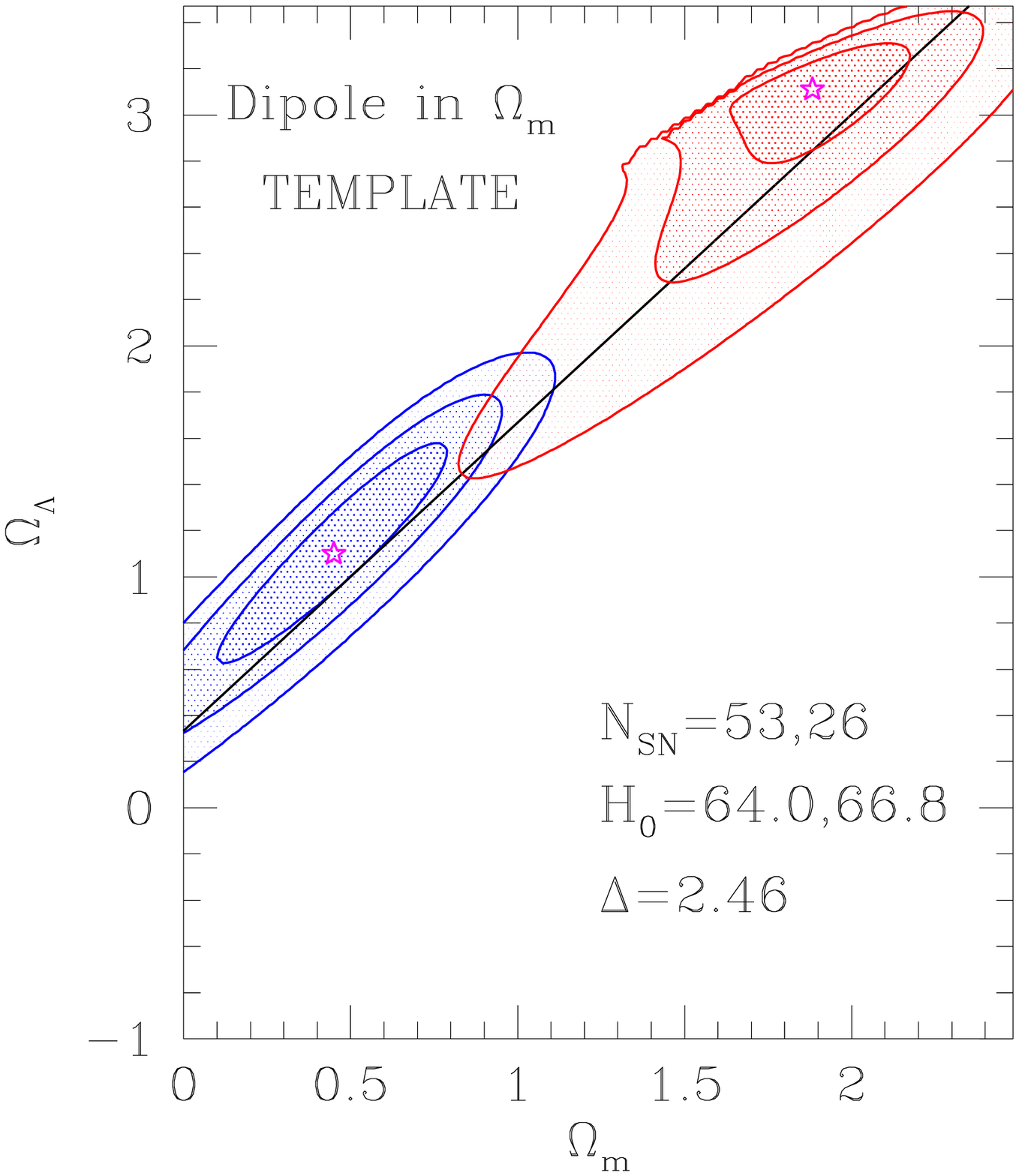}}
\vskip1truecm
{\epsfxsize=2.7 in \epsfbox{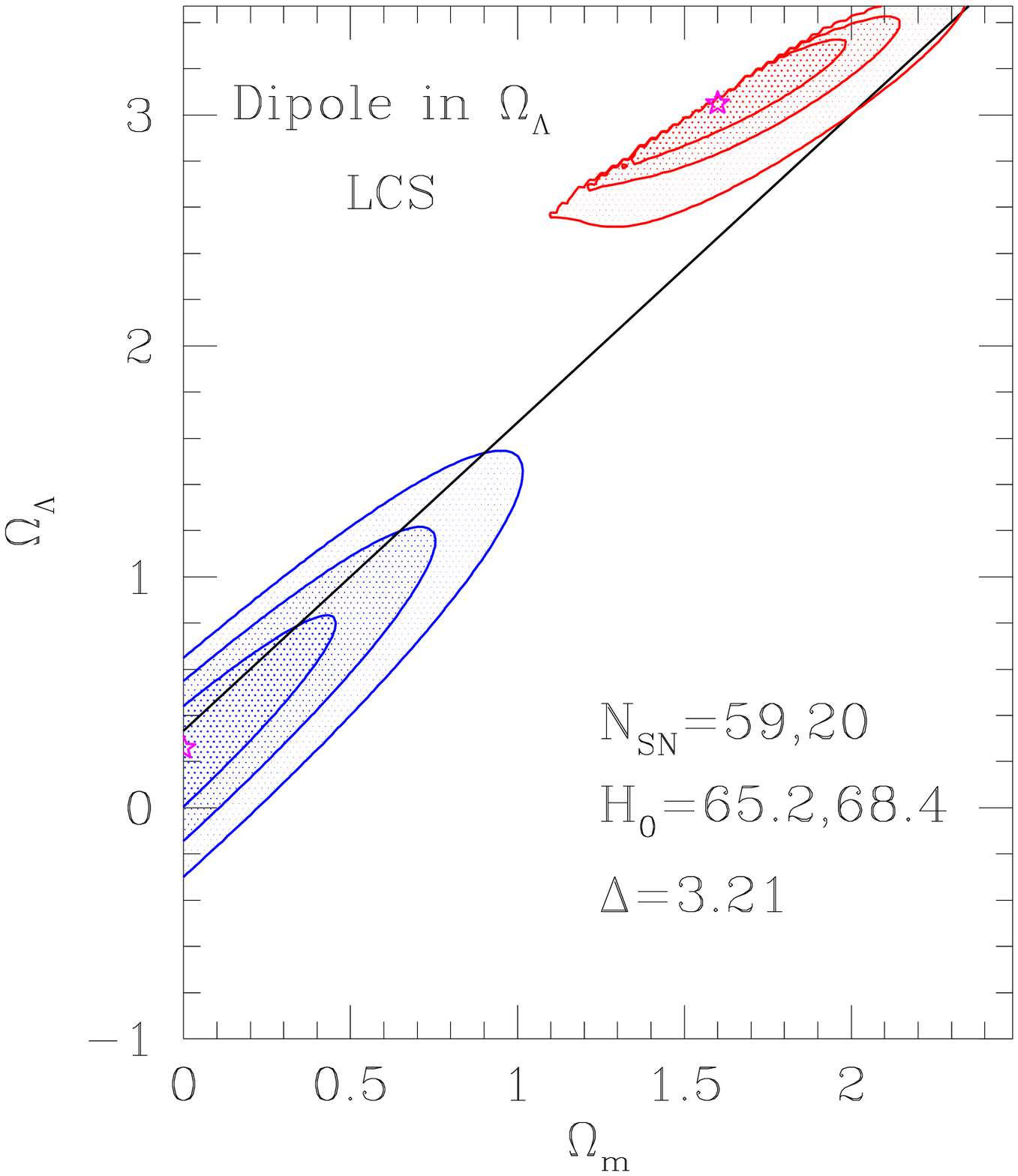}}
{\epsfxsize=2.7 in \epsfbox{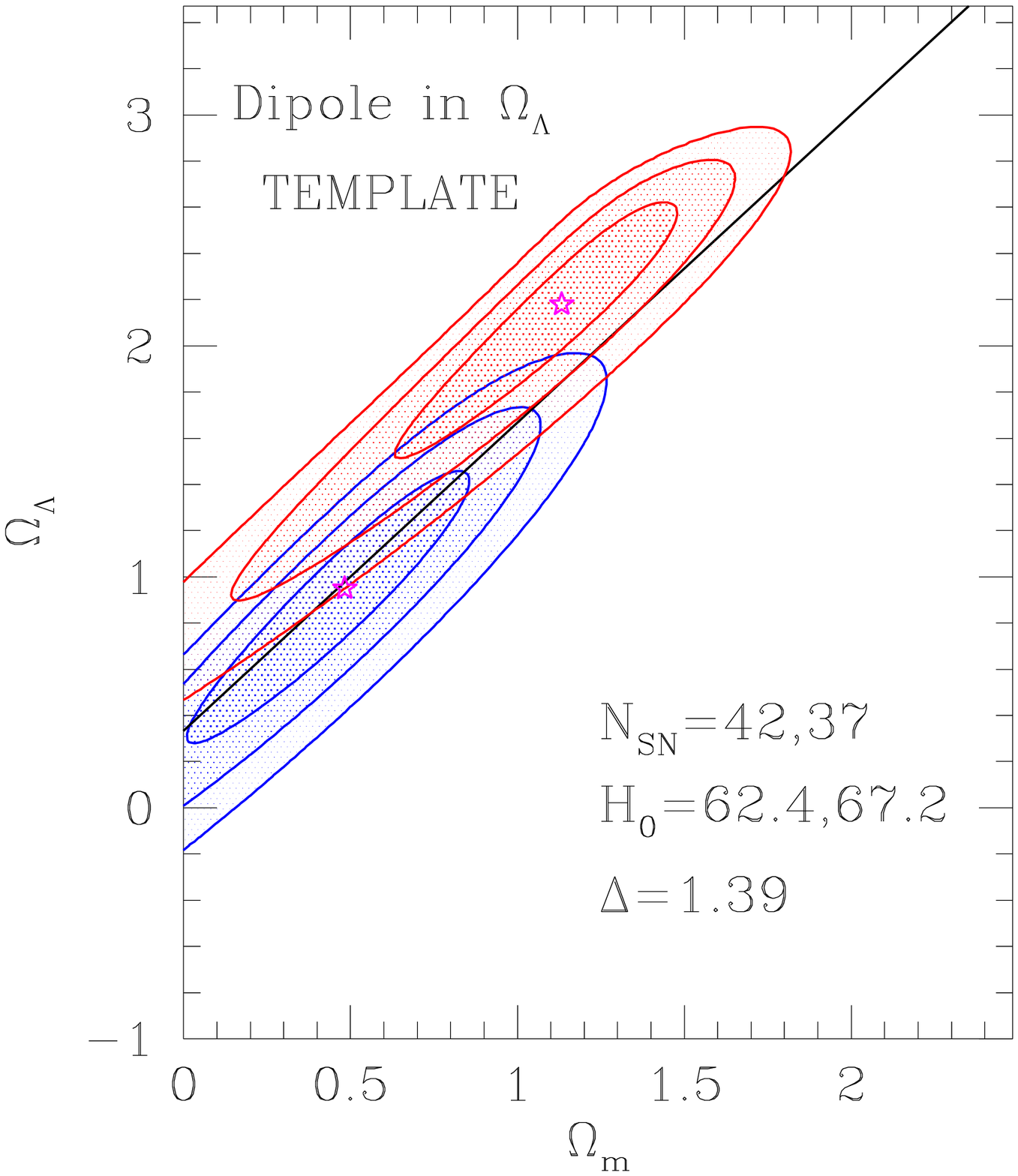}}
\caption{
Confidence regions (same as Fig. \ref{fig:xlikeA} 
drawn from 79 SNe of the unified sample in two
hemispheres that maximize the $\Omega_M$ dipole (up) and
$\Omega_\Lambda$ dipole (bottom) using the
LCS method (left) and the TEMPLATE method (right).
Marked are the number of SNe in each hemisphere, the best-fit $H_0$, 
and the distance between peak probabilities ($\Delta$).
}
\label{fig:sn_2con}
\end{figure*}

Figure \ref{fig:sn_2con} verify the fact that the current SNe data
best constrain a linear combination of the cosmological parameters
$\Omega_M$, $\Omega_\Lambda$. In all four panels the likelihood maxima
move along the line (P99) $-4\Omega_M +3 \Omega_{\Lambda} = 1$,
sometimes with a large distance $\Delta$ between the two maxima for the
two disjoint hemispheres (quoted on the plots).
In three cases the contour levels overlap significantly (see next
section for quantitative evaluation). In the case of the
$\Omega_\Lambda$ dipole, using the LCS method, there is no overlap between
the $99.7\%$ confidence levels of the two hemispheres.
The discrepancy stems from the very assymetric distribution of SNe
between the two hemispheres (59 on one versus 20 on the other) and only
one SN (1995at) with $z>0.5$ in the 20 SNe sample. A small error in the
distance measurement of this SN, or a systematic deviation of it from
the average LCS relation may cause such a discrepancy as we demonstrate
in the next section. 
Elimination of this SN yields a dipole which points $\sim20\deg$ away
from the original direction, reduced $\Delta$ value of $2.39$, and
almost full inclusion of the $99.7\%$ confidence contour for the larger
sample ($56$ SNe)
within the $95.4\%$ confidence level of the remaining $22$ SNe.

Note that a different ``mixture'' of redshift
distribution to different directions may cause some directions to become
more sensitive to one parameter. E.g, SNe at $z\simeq0.4-0.5$ are mostly
sensitive to the $\Omega_M - \Omega_\Lambda$ combination, as opposed to
higher weight on $\Omega_M$ as redshift increases
($\vert \partial\Delta/\partial \Omega_M\vert  
  > \vert \partial \Delta / \partial \Omega_\Lambda\vert $).
Figure \ref{fig:SN_sample} includes the dipole directions (positive)
of $\Omega$ in both methods. We observe no coincidence with any Galactic
or CMB direction, moreover not all dipoles point to the same direction.

The dipole of the $\delta_{\chi^2_{\rm HD}}$ field points in both
methods toward $(l=80\deg,b=-20\deg)$ with $\langle \chi^2_{\rm HD} \rangle
= 1.00\, (1.01)$ and largest difference of $0.60\,(0.63)$ for the LCS
(TEMPLATE) method. This dipole direction is suspiciously close to the
Galactic plane.

One worry is that the detected signal is due to the (mis)match
between the two data sets. We therefore repeated the computation for
each
data set separately and verified that though the noise level increases,
the results as drawn from each one of the data sets are
consistent with the results from the unified set both in magnitude and
direction.

\section{Degree of anisotropy}
\label{sec:degree}

The results of the last section, regarding the spherical harmonic
expansion and the various dipole magnitudes, should now be put in an
expected distribution in order to draw conclusions about the degree of
anisotropy.

The hypothesis we are trying to address is that the SN data do
not falsify the FRW geometry as a reliable description of the $z\simeq1$
Universe. This strategy is more efficient than addressing specific
anisotropic cosmological models \cite{celerier:00,celerier:00a}.
We therefore compute the probability distribution of the dipole
magnitudes within a FRW universe and confront it with the values
obtained for the real Universe. 

A simple two dimensional Kolmogorov-Smirnov test to falsify the
hypothesis that the two contour maps come from the same underlying
distribution of cosmological parameters is inadequate here. Since we have
used the maximum discrepant values in order to obtain the dipole, the
two sub-samples are not randomly selected and therefore can not be 
confronted in a KS test.

The probability distribution depends on the actual cosmological values
and to a lesser extent on the power spectrum (via the scatter due
to potential fluctuations). For a self consistency check, the underlying
cosmology is taken to be the "best fit"
cosmological model (\S\ref{sec:unified}), which we then sample by Monte-Carlo simulations.

To mimic accurately the SN sample, we use the same angular locations and
redshift values as of the observed sample. Luminosity distances,
magnitude scatter and peculiar velocities are drawn from Gaussian
distributions with the appropriate observed standard deviation.

The dipole analysis is repeated for $200$ mock catalogs of the SN and the
maximal dipole magnitude is calculated to obtain its distribution for
{\it the current sampled} SNe.

Table 1
shows the rejection levels of
the hypothesis that the Universe up to $z\simeq1$ can be described by
a FRW metric. E.g., using the LCS method and the current sample of SNIa
we expect in $19\%$ of all cases to detect a higher $\Delta$ value 
for $\Omega_M$ dipole, than the observed one.

\vskip0.5truecm

%\hskip-2truecm
\begin{tabular}{|c||c|c|}
\multicolumn{3}{|c|}{Table 1} \\
\hline
\multicolumn{3}{|c|}{Isotropy rejection levels using $\Delta$} \\
\hline\hline
%first line
\multicolumn{1}{|c||}{Cosmological}   
&\multicolumn{2}{c|}{Method} \\ \cline{2-3}
%second line
\multicolumn{1}{|c||}{parameter}
&\multicolumn{1}{c|}{LCS} 
&\multicolumn{1}{c|}{TEMPLATE} \\ \hline\hline
%%%
\multicolumn{1}{|c||}{$H_0$}
&\multicolumn{1}{c|}{33\%} 
&\multicolumn{1}{c|}{70\%} \\ \cline{1-3} 
%%%
\multicolumn{1}{|c||}{$\Omega_M$}
&\multicolumn{1}{c|}{81\%} 
&\multicolumn{1}{c|}{79\%} \\ \cline{1-3} 
%%%
\multicolumn{1}{|c||}{$\Omega_\Lambda$}
&\multicolumn{1}{c|}{88\%} 
&\multicolumn{1}{c|}{64\%} \\ \cline{1-3} 
\end{tabular}

\section {Discussion} 
\label{sec:discuss}

By the exploitation of the current available SNe data we have put
constraints on the rejection level of the cosmological principle
validity up to $z\simeq 1$. A FRW metric is found to be an adequate
description of the Universe. In $\sim 20\%$ of all 
realizations of such
universes, the dipole signature for anisotropy in the cosmological 
parameters $H_0$, $\Omega_M$ and $\Omega_\Lambda$ exceeds the 
observed one.

Even though such dipole magnitudes are reasonable in the framework of the 
FRW model, they may be indicative of non-cosmological contributions to
the angular power spectrum.
In \S\ref{sec:intro} we listed possible such contributions. 
If indeed the Universe up to $z\simeq 1$ is well represented by a FRW
metric then we can exclude large coherent structures at $z \gsim
0.3$. Such are the structures that may lead to dipole and quadrupole 
signatures due to coherent gravitational lensing 
magnification/de-magnification and therefore the latter can be excluded
as anisotropy contributors. 

That leaves small scale (Galactic) foreground effects to be the most
likely power contributors. 
The Galactic disk geometry makes the 
quadrupole the most significant multipole to be considered, though the
solar system offset from the Galactic center may bring about a dipole
contribution as well.
In the current sample the quadrupole term is only slightly larger
than the noise level and does not allow any conclusive results.
None of the dipole directions for $\Omega$ coincides with the Galactic
plane and thus they are probably not correlated with it. 
Multipoles due to dust extinction may be affirmed by multiple expansion 
of the residual colors after extinction correction (i.e., R98 and P99 
appendices).

The one case where two significantly non-overlapping confidence regions
are found for two hemispheres that maximize the
$\Omega_\Lambda$ dipole (LCS), is probably due to a single SN (1995at)
for which the individual errors have been underestimated. This case is
an exception since the overall $\chi^2$ values for the HD fits 
are statistically acceptable.
Nevertheless, this case demonstrates the hazard in the draw of conclusions
based on a handful of SNe, for which the error in the error estimate is
uncertain.

In general, the TEMPLATE method provides a better statistical agreement
of the data with an FRW model and the current SNe data. This is seen
from the magnitude match (cf. \S\ref{sec:data}), tighter
constraints from the combined set, smaller noise levels for all
multipoles, and smaller $\Delta$ values for the $\Omega$ dipoles.

We conclude that an isotropic $z\sim1$ universe cannot be rejected by
more than a $1\sigma$ level based on the current SNe data.

\section*{Acknowledgments:} 
This work was supported by the US-Israel Binational Science Foundation,
by the Israel Science Foundation, and by grants from NASA and NSF at UCSC.

\bibliographystyle{mnras}
\bibliography{mnrasmnemonic,refs}
\end{document}